\documentclass{ws-procs9x6}

\begin{document}

\title{On theory of Landau-Pomeranchuk-Migdal effect
\footnote{\uppercase{T}his work is partially supported by grant
03-02-16154 of the \uppercase{R}ussian \uppercase{F}und of
\uppercase{F}undamental \uppercase{R}esearch}}

\author{V.~N.BAIER and V.~M.KATKOV}

\address{Budker Institute of Nuclear Physics, \\
Novosibirsk, 630090, Russia \\
E-mail: baier@inp.nsk.su; katkov@inp.nsk.su}


\maketitle

\abstracts{The cross section of bremsstrahlung from high-energy
electron suppressed due to the multiple scattering of an emitting
electron in a dense media (the LPM effect) is discussed.
Development of the LPM effect theory is outlined. For the
target of finite thickness one has to include boundary photon
radiation as well as the interference effects. The importance of 
multiphoton effects is emphasized. Comparison of theory with 
SLAC data is presented.}

\section{Introduction}

The process of photon radiation takes place not in one point but
in some domain of space-time, in which a photon (an emitted wave
in the classical language) is originating. The longitudinal
dimension of this domain is called {\it the formation (coherence)
length} $l_f$
\begin{equation}
l_{f0}(\omega)=2\varepsilon \varepsilon'/\omega m^2, \label{6}
\end{equation}
where $\varepsilon (\varepsilon')$ is the energy of the initial
(final) electron, $\omega$ is the photon energy,
$\varepsilon'=\varepsilon-\omega$, $m$ is the electron mass, we
employ units such that $\hbar=c=1$. So for ultrarelativistic
particles the formation length extends substantially. For example
for $\varepsilon=25$~GeV and emission of $\omega=100$~MeV photon,
$l_{f}=10\mu$m, i.e.$\sim 10^5$ interatomic distances.

Landau and Pomeranchuk were the first to show that if the
formation length of the bremsstrahlung becomes comparable to the
distance over which the multiple scattering becomes important,
the bremsstrahlung will be suppressed \cite{LP}. They considered
radiation of soft photons. Migdal \cite{M} developed a
quantitative theory of this phenomenon. Now the common name is
the Landau- Pomeranchuk -Migdal (LPM) effect.

Let us estimate a disturbance of the emission process due to a
multiple scattering. As it is known, the mean angle of the
multiple scattering at some length $l_{f}$ is
\begin{equation}
\vartheta_s = \sqrt{\vartheta_s^2} = \varepsilon_s/\varepsilon
\sqrt{l_{f}/2L_{rad}},\quad
\varepsilon_s=m\sqrt{4\pi/\alpha}=21.2~{\rm MeV}, \label{8}
\end{equation}
where $\alpha=e^2=1/137$, $L_{rad}$ is the radiation length.
Since we are interesting in influence of the multiple scattering
on the radiation process we put here the formation length
(\ref{6}). One can expect that when $\vartheta_s \geq 1/\gamma$
this influence will be substantial. From this inequality we have
\begin{equation}
\varepsilon'/\varepsilon_{LP} \geq \omega/\varepsilon, \quad
\varepsilon_{LP}=m^4 L_{rad}/\varepsilon_s^2, \label{9}
\end{equation}
here $\varepsilon_{LP}$ is the characteristic energy scale, for
which the multiple scattering will influence the radiation
process for the whole spectrum. It was introduced in \cite{LP} and
denoted by $E_0=\varepsilon_{LP}$. For tungsten we have
$\varepsilon_{LP}=2.72$~TeV, and similar values for the all heavy
elements. For light elements the energy $\varepsilon_{LP}$ is much
larger. When $\varepsilon \ll \varepsilon_{LP}$ the multiple
scattering will influence emission of soft photons only with
energy $\omega \leq \omega_{LP}=\varepsilon^2/\varepsilon_{LP}$,
e.g. for particles with energy $\varepsilon=25$~GeV one has
$\omega_{LP} \simeq 230$~MeV. Of course, we estimate here the
criterion only. For description of the effect one has to calculate
the probability of the bremsstrahlung taking into account multiple
scattering.

\section{The LPM effect in a infinite medium}

We consider first the case where the formation length is much
shorter than the thickness of a target $l (l_{f0} \ll l)$. The
basic formulas for probability of radiation under influence of
multiple scattering were derived in \cite{BKS1}. The general
expression for the radiation probability obtained in framework of
the quasiclassical operator method (Eq.(4.2) in \cite{BKS2})was
averaged over all possible particle trajectories. This operation
was performed with the aid of the distribution function, averaged
over atomic positions in the scattering medium and satisfying the
kinetic equation. The spectral distribution of the probability of
radiation per unit time was reduced in \cite{BK1} to the form
\begin{equation}
dW/d\omega=(2\alpha/\gamma^2) {\rm Re} \int_{0}^{\infty} dt
e^{-it}\left[R_1 \varphi_0(0, t) + R_2 {\bf p}
\mbox{\boldmath$\varphi$}(0, t)\right], \label{2.8}
\end{equation}
where $R_1=\omega^2/\varepsilon \varepsilon',\quad R_2
=\varepsilon/\varepsilon'+\varepsilon'/\varepsilon$, and the
functions $\varphi_{\mu}~(\varphi_{0},
\mbox{\boldmath$\varphi$})$ satisfy the equation
\begin{eqnarray}
&&\displaystyle{i\frac{\partial \varphi_{\mu}}{\partial t}= H
\varphi_{\mu},\quad H={\bf p}^2 -
iV(\mbox{\boldmath$\varrho$}),\quad {\bf p}= -i
\mbox{\boldmath$\nabla$}_{\mbox{\boldmath$\varrho$}},\quad
V(\mbox{\boldmath$\varrho$})=-Q\mbox{\boldmath$\varrho$}^2
\Big(\ln \gamma^2 \vartheta_1^2}
\nonumber \\
&&\displaystyle{+\ln
\frac{\mbox{\boldmath$\varrho$}^2}{4}+2C-1\Big),\quad
Q=\frac{2\pi n_a Z^2 \alpha^2 \varepsilon \varepsilon'}{m^4
\omega},\quad C=0.577216...} \label{2.9}
\end{eqnarray}
with the initial conditions $\varphi_0(\mbox{\boldmath$\varrho$},
0)= \delta(\mbox{\boldmath$\varrho$}), \quad
\mbox{\boldmath$\varphi$}(\mbox{\boldmath$\varrho$}, 0) ={\bf p}
\delta(\mbox{\boldmath$\varrho$})$. Here $\varepsilon,
\varepsilon', \omega$ are defined in (\ref{6}) $Z$ is the charge
of the nucleus and $n_a$ is the number density of atoms in the
medium, $\displaystyle{\vartheta_1=1/\varepsilon a_s}$, $a_s$ is
the screening radius ($a_s=0.81a_BZ^{-1/3}$, $a_B$ is the Bohr
radius). Note, that it is implied that in formula (\ref{2.8})
subtraction at $V=0$ is made.

The potential $V(\mbox{\boldmath$\varrho$})$ (\ref{2.9})
corresponds to consideration of scattering in the Born
approximation. The difference of exact as a function of
$Z\alpha$  potential $V(\mbox{\boldmath$\varrho$})$ and taken in
the Born approximation was computed in Appendix A of \cite{BK1}.
The potential $V(\mbox{\boldmath$\varrho$})$ with the Coulomb
corrections taken into account is
\begin{equation}
V(\mbox{\boldmath$\varrho$})=-Q\mbox{\boldmath$\varrho$}^2
\left(\ln \gamma^2 \vartheta_2^2 +\ln
\mbox{\boldmath$\varrho$}^2/4+2C\right),
\label{2.9a}
\end{equation}
where $\vartheta_2=\vartheta_1 \exp (f-1/2)$, the function
$f=f(Z\alpha)$ is
\begin{equation}
f(\xi)={\rm Re}\left[\psi(1+i\xi)-\psi(1) \right]=
\xi^2\sum_{n=1}^{\infty}1/(n(n^2+\xi^2)),
\label{2.9b}
\end{equation}
where $\psi(\xi)$ is the logarithmic derivative of the gamma
function.

In above formulas $\mbox{\boldmath$\varrho$}$ is two-dimensional
space of the impact parameters measured in the Compton
wavelengths $\lambda_c$, which is conjugate to space of the
transverse momentum transfers measured in the electron mass $m$.

An operator form of a solution of Eq. (\ref{2.9}) is
\begin{eqnarray}
&&\varphi_0(\mbox{\boldmath$\varrho$}, t) = \exp(-iHt)
\varphi_0(\mbox{\boldmath$\varrho$}, 0)=
<\mbox{\boldmath$\varrho$}| \exp(-iHt) |0>,\quad  H={\bf
p}^2-iV(\mbox{\boldmath$\varrho$}),
\nonumber \\
&&\mbox{\boldmath$\varphi$}(\mbox{\boldmath$\varrho$}, t)=
\exp(-iHt) {\bf p} \varphi_0(\mbox{\boldmath$\varrho$},0)=
<\mbox{\boldmath$\varrho$}| \exp(-iHt){\bf p} |0>, \label{2.10}
\end{eqnarray}
where we introduce the following Dirac state vectors:
$|\mbox{\boldmath$\varrho$}>$ is the state vector of coordinate
$\mbox{\boldmath$\varrho$}$, and
$<\mbox{\boldmath$\varrho$}|0>=\delta(\mbox{\boldmath$\varrho$})$.
Substituting (\ref{2.10}) into (\ref{2.8}) and taking integral
over $t$ we obtain for the spectral distribution of the
probability of radiation
\begin{equation}
\frac{dW}{d\omega}=\frac{2\alpha}{\gamma^2} {\rm Im} T,\quad T=
<0| R_1 \left(G^{-1}-G_0^{-1} \right) + R_2 {\bf p}
\left(G^{-1}-G_0^{-1} \right) {\bf p} |0>, \label{2.11}
\end{equation}
where
\begin{equation}
G={\bf p}^2+1-iV,\quad G_0={\bf p}^2+1. \label{2.12}
\end{equation}
Here and below we consider an expression $<0|...|0>$ as a limit:
${\rm lim}~{\bf x} \rightarrow 0,
\newline {\rm lim}~{\bf x'} \rightarrow 0$ of
$<{\bf x}|...|{\bf x'}>$.

Now we estimate the effective impact parameters $\varrho_c$ which
give the main contribution into the radiation probability. Since
the characteristic values of $\varrho_c$ can be found
straightforwardly by calculation of (\ref{2.11}), we the estimate
characteristic angles $\vartheta_c$ connected with $\varrho_c$ by
an equality $\varrho_c=1/(\gamma \vartheta_c)$. The mean square
scattering angle of a particle on the formation length of a
photon $l_f$ (\ref{6}) has the form
\begin{equation}
\vartheta_s^2=\frac{4\pi Z^2 \alpha^2}{\varepsilon^2}n_a l_f \ln
\frac{\zeta}{\gamma^2 \vartheta_1^2}= \frac{4Q}{\gamma^2\zeta}
\ln \frac{\zeta}{\gamma^2 \vartheta_1^2}, \label{2.13}
\end{equation}
where $\zeta=1+\gamma^2\vartheta^2$, we neglect here the
polarization of a medium. When $\vartheta_s^2 \ll 1/\gamma^2$ the
contribution in the probability of radiation gives a region where
$\zeta \sim 1 (\vartheta_c=1/\gamma)$, in this case
$\varrho_c=1$. When $\vartheta_s \gg 1/\gamma$ the characteristic
angle of radiation is determined by  self-consistency arguments:
\begin{equation}
\vartheta_s^2 \simeq \vartheta_c^2 \simeq
\frac{\zeta_c}{\gamma^2}= \frac{4Q}{\zeta_c \gamma^2} \ln
\frac{\zeta_c}{\gamma^2\vartheta_1^2},\quad \frac{4Q}{\zeta_c^2}
\ln \frac{\zeta_c}{\gamma^2\vartheta_1^2}=1,\quad 4Q\varrho_c^4
\ln \frac{1}{\gamma^2\vartheta_1^2 \varrho_c^2}=1. \label{2.14}
\end{equation}
It should be noted that when the characteristic impact parameter
$\varrho_c$ becomes smaller than the radius of nucleus $R_n$, the
potential $V(\mbox{\boldmath$\varrho$})$ acquires an oscillator
form (see Appendix B, Eq.(B.3) in \cite{BK1})
\begin{equation}
V(\mbox{\boldmath$\varrho$})=Q\mbox{\boldmath$\varrho$}^2
\left(\ln a_s^2/R_n^2-0.041 \right) \label{2.15}
\end{equation}

Allowing for the estimates (\ref{2.14}) we present the potential
$V(\mbox{\boldmath$\varrho$})$ (\ref{2.9}) in the following form
\begin{eqnarray}
&&\displaystyle{V(\mbox{\boldmath$\varrho$})=V_c(\mbox{\boldmath$\varrho$})+
v(\mbox{\boldmath$\varrho$}),~ V_c(\mbox{\boldmath$\varrho$})=
q\mbox{\boldmath$\varrho$}^2,~ q=QL_c,\quad L_c \equiv
L(\varrho_c)=\ln \frac{1}{\gamma^2\vartheta_2^2\varrho_c^2}},
\nonumber \\
&&\displaystyle{L_1 \equiv L(1)=\ln
\frac{1}{\gamma^2\vartheta_2^2}= \ln
\frac{a_{s2}^2}{\lambda_c^2}, \quad v(\mbox{\boldmath$\varrho$})=
-\frac{q\mbox{\boldmath$\varrho$}^2}{L_c} \left(2C+\ln
\frac{\mbox{\boldmath$\varrho$}^2}{4\varrho_c^2} \right)}.
\label{2.16}
\end{eqnarray}
The inclusion of the Coulomb corrections ($f(Z\alpha)$ and -1)
into $\ln \vartheta_2^2$ diminishes effectively the correction
$v(\mbox{\boldmath$\varrho$})$ to the potential
$V_c(\mbox{\boldmath$\varrho$})$. In accordance with such
division of the potential we present the propagators in
expression (\ref{2.11}) as
\begin{equation}
G^{-1}-G_0^{-1}=G^{-1}-G_c^{-1} + G_c^{-1}-G_0^{-1} \label{2.17}
\end{equation}
where
\[
G_c={\bf p}^2+1-iV_c,\quad G={\bf p}^2+1-iV_c-iv
\]
This representation of the propagator $G^{-1}$ permits one to
expand it over the "perturbation" $v$. Indeed, with an increase of
$q$ the relative value of the perturbation is diminished
$\displaystyle{(\frac{v}{V_c} \sim \frac{1}{L_c})}$ since the
effective impact parameters diminishes and, correspondingly, the
value of logarithm $L_c$ in (\ref{2.16}) increases. The maximal
value of $L_c$ is determined by the size of a nucleus $R_n$
\begin{equation}
L_{max}=\ln \frac{a_{s2}^2}{R_n^2} \simeq 2 \ln
\frac{a_{s2}^2}{\lambda_c^2} \equiv 2 L_1,\quad L_1=2(\ln
(183Z^{-1/3})-f(Z\alpha)), \label{2.18}
\end{equation}
where $a_{s2}=a_s \exp (-f+1/2)$. So, one can to redefine the
parameters $a_s$ and $\vartheta_1$ to include the Coulomb
corrections. The value $L_1$ is the important parameter of the
radiation theory.

The matrix elements of the operator $G_c^{-1}$ can be calculated
explicitly. The exponential parametrization of the propagator is
\begin{equation}
\displaystyle{G_c^{-1}=i\int_{0}^{\infty}dt e^{-it} \exp (-i H_c
t),\quad H_c={\bf p}^2 - iq \mbox{\boldmath$\varrho$}^2}
\label{2.19}
\end{equation}
The matrix elements of the operator $\displaystyle{\exp (-i H_c
t)}$ has the form (details see in \cite{BK1})
\begin{eqnarray}
&& <\mbox{\boldmath$\varrho$}_1|\exp(-i
H_ct)|\mbox{\boldmath$\varrho$}_2> \equiv
K_c(\mbox{\boldmath$\varrho$}_1, \mbox{\boldmath$\varrho$}_2,
t),\quad K_c(\mbox{\boldmath$\varrho$}_1,
\mbox{\boldmath$\varrho$}_2, t)
\nonumber \\
&& = \frac{\nu}{4\pi i \sinh \nu t} \exp \left\{ \frac{i\nu}{4}
\left[
(\mbox{\boldmath$\varrho$}_1^2+\mbox{\boldmath$\varrho$}_2^2)
\coth \nu t - \frac{2}{\sinh \nu t}
\mbox{\boldmath$\varrho$}_1\mbox{\boldmath$\varrho$}_2\right]
\right\}, \label{2.24}
\end{eqnarray}
where $\nu=2\sqrt{iq}$ (see (\ref{2.16})).

Substituting formulae (\ref{2.19}) and (\ref{2.24}) in the
expression for the spectral distribution of the probability of
radiation (\ref{2.11}) we have
\begin{eqnarray}
\hspace{-13mm}&&\displaystyle{\frac{dW_c}{d\omega}=
\frac{\alpha}{2\pi \gamma^2} {\rm Im}~\Phi (\nu)},
\nonumber \\
\hspace{-13mm}&&\displaystyle{\Phi(\nu)=\nu\int_{0}^{\infty} dt
e^{-it}\left[R_1 \left(1/\sinh z-1/z\right)-i\nu R_2 \left(
1/\sinh^2z- 1/z^2\right) \right]}
\nonumber \\
\hspace{-13mm}&&=R_1\left(\ln p -\psi\left(p+1/2 \right) \right)
+R_2\left( \psi\left(p \right)-\ln p+1/2p\right) \label{2.25}
\end{eqnarray}
where $z=\nu t,~p=i/(2\nu)$, let us remind that $\psi(x)$ is the
logarithmic derivative of the gamma function (see
Eq.(\ref{2.9b})). If in Eq.(\ref{2.25}) one omits the Coulomb
correction, then the probability (\ref{2.25}) coincides formally
with the probability calculated by Migdal (see Eq.(49) in
\cite{M}).

We now expand the expression $G^{-1}-G_c^{-1}$ over powers of $v$
\begin{equation}
G^{-1}-G_c^{-1}=G_c^{-1}(iv)G_c^{-1}+G_c^{-1}(iv)G_c^{-1}(iv)G_c^{-1}
+... \label{2.26}
\end{equation}

In accordance with (\ref{2.17}) and (\ref{2.26}) we present the
probability of radiation in the form
\begin{equation}
dW/d\omega=dW_c/d\omega+dW_1/d\omega+ dW_2/d\omega+...
\label{2.27}
\end{equation}
At $Q \geq 1$ the expansion (\ref{2.26}) is a series over powers
of $\displaystyle{1/L}$. It is important that variation of the
parameter $\varrho_c$ by a factor order of 1 has an influence on
the dropped terms in (\ref{2.26}) only. The probability of
radiation $\displaystyle{dW_c/d\omega}$ is defined by
Eq.(\ref{2.25}).

The term $\displaystyle{dW_1/d\omega}$ in (\ref{2.27})
corresponds to the term linear in $v$ in (\ref{2.26}). The
explicit formula for the first correction to the probability of
radiation \cite{BK1} is
\begin{eqnarray}
\hspace{-4mm}&&\displaystyle{\frac{dW_1}{d\omega}=-\frac{\alpha}{4\pi
\gamma^2 L_c} {\rm Im}~F(\nu);\quad -{\rm
Im}~F(\nu)=D_1(\nu_0)R_1+\frac{1}{s}D_2(\nu_0)R_2};
\nonumber \\
\hspace{-4mm}&&\displaystyle{D_1(\nu_0)=\int_{0}^{\infty}\frac{dz
e^{-sz}}{\sinh^2z} \left[d(z)\sin sz+\frac{\pi}{4}g(z)\cos sz
\right],\quad D_2(\nu_0)=\int_{0}^{\infty}\frac{dz
e^{-sz}}{\sinh^3 z}}
\nonumber \\
\hspace{-4mm}&&\displaystyle{\times \left\{\left[d(z)-
\frac{1}{2}g(z)\right]\left(\sin sz+\cos sz\right)
+\frac{\pi}{4}g(z)\left(\cos sz - \sin sz\right)\right\}},
\nonumber \\
\hspace{-4mm}&&\displaystyle{d(z)=(\ln \nu_0
\vartheta(1-\nu_0)-\ln \sinh z -C)g(z) -2\cosh z
G(z),~s=1/\sqrt{2}\nu_0}, \label{2.30}
\end{eqnarray}
where
\begin{eqnarray}
&&g(z)=z \cosh z-\sinh z,\quad G(z)=\int_{0}^{z}(1-y\coth y)dy
\nonumber \\
&&=z-z^2/2-\pi^2/12- z\ln \left(1-e^{-2z} \right) +{\rm Li}_2
\left(e^{-2z}\right)/2, \label{2.31a}
\end{eqnarray}
here ${\rm Li}_2 \left(x \right)$ is the Euler dilogarithm; and
\begin{equation}
\nu_0^2 \equiv |\nu|^2=4q= 4 QL(\varrho_c)= 8\pi n_a Z^2\alpha^2
\varepsilon \varepsilon'L(\varrho_c)/m^4 \omega, \label{2.32a}
\end{equation}

As it was said above (see (\ref{2.14}), (\ref{2.18})),
$\varrho_c=1$ at
\begin{equation}
|\nu^2|=\nu_1^2=4QL_1 \leq 1. \label{2.30a}
\end{equation}
The logarithmic functions $L_c \equiv L(\varrho_c)$ and $L_1$ are
defined in (\ref{2.16}) and (\ref{2.18}). If the parameter $|\nu|
> 1$, the value of $\varrho_c$ is defined from the equation
(\ref{2.14}), where $\vartheta_1 \rightarrow \vartheta_2$, up to
$\varrho_c=R_n/\lambda_c$.  So, we have two representation of
$|\nu|$ depending on $\varrho_c$: at $\varrho_c = 1$ it is
$|\nu|=\nu_1$ and at $\varrho_c \leq 1$ it is $|\nu|=\nu_0$. The
mentioned parameters can be presented in the following form
\begin{eqnarray}
\hspace{-8mm}&& \nu^2=i\nu_0^2,\quad \nu_0^2=|\nu|^2 \simeq
\nu_1^2\left(1+ \frac{\ln \nu_1}{L_1}\vartheta(\nu_1-1)
\right),\quad
\nu_1^2=\frac{\varepsilon}{\varepsilon_e}\frac{1-x}{x},
\nonumber \\
\hspace{-8mm}&&\varepsilon_e=m\left(8\pi Z^2 \alpha^2 n_a
\lambda_c^3 L_1 \right)^{-1},~ L_c \simeq L_1 \left(1+ \frac{\ln
\nu_1}{L_1}\vartheta(\nu_1-1)
\right),~x=\frac{\omega}{\varepsilon}. \label{2.30b}
\end{eqnarray}
It should be noted that in the logarithmic approximation the
parameter $\varrho_c$ entering into the parameter $\nu$ is
defined up to the factor $\sim 1$. However, we calculated the
next term of the decomposition over
$v(\mbox{\boldmath$\varrho$})$ (an accuracy up to the "next to
leading logarithm") and this permitted to obtain the result which
is independent of the parameter $\varrho_c$ (in terms $\propto
1/L$). Our definition of the parameter $\varrho_c$ minimizes
corrections to (\ref{2.25}) practically for all values of the
parameter $\varrho_c$.

The approximate solution of Eq.(\ref{2.14}) given in this formula
has quite good numerical accuracy:it is $\sim 2~\%$ at $\nu_1=100$
and $\sim 4.5~\%$ at $\nu_1=1000$.The LPM effect manifests itself
when
\begin{equation}
\nu_1(x_c)=1,\quad
x_c=\omega_c/\varepsilon=\varepsilon/(\varepsilon_e+\varepsilon).
\label{2.31b}
\end{equation}
So, the characteristic energy $\varepsilon_e$ is the energy
starting from which the multiple scattering distorts the whole
spectrum of radiation including its hard part. If the radiation
length $L_{rad}$ is taken within the logarithmic approximation
the value $\varepsilon_e$ coincides with $\varepsilon_{LP}$
Eq.(\ref{9}). The formulas derived in \cite{BK1},\cite{L5} and
written down above are valid for any energy. In Fig.1 the spectral
radiation intensity in gold ($\varepsilon_e=2.5$~TeV) is shown
for different energies of the initial electron. In the case when
$\varepsilon \ll \varepsilon_e$ ($\varepsilon=25~$GeV and
$\varepsilon=250~$GeV) the LPM suppression is seen in the soft
part of the spectrum only for $x \leq x_c \simeq
\varepsilon/\varepsilon_e \ll 1$ while in the region $\varepsilon
\geq \varepsilon_e$ ($\varepsilon=2.5~$TeV and
$\varepsilon=25~$TeV) where $x_c \sim 1$ the LPM effect is
significant  for any $x$.

\begin{figure}[ht]
\centerline{\epsfxsize=9.6cm\epsfbox{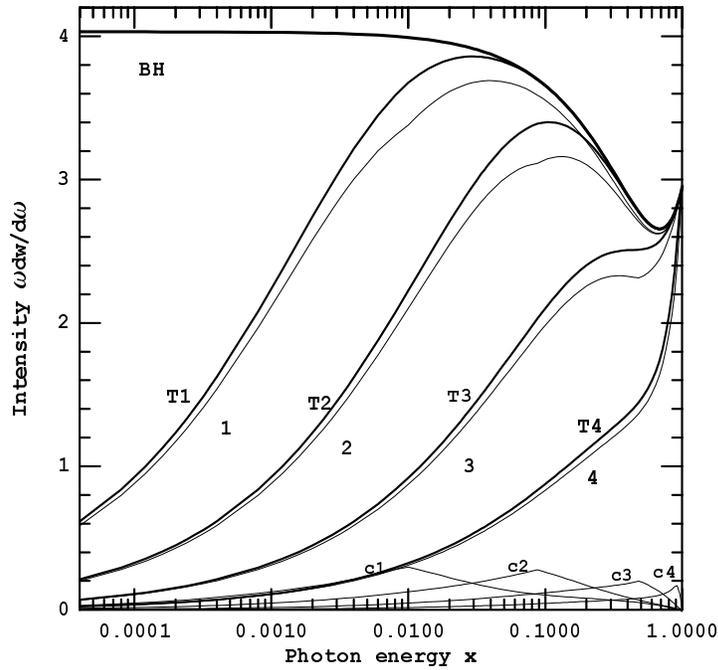}} \caption{The
spectral intensity of radiation $\omega dW/d\omega= xdW/dx,~x=
\omega/\varepsilon$ in gold in terms of $3 L_{rad}$ taken with
the Coulomb corrections.
 Curve BH is the Bethe-Maximon spectral intensity
(see Eq.(\ref{2.35}));
 curve 1 is the logarithmic approximation
$\omega dW_c/d\omega$ Eq.(\ref{2.25}), curve c1 is the first
correction to the spectral intensity $\omega dW_1/d\omega$
Eq.(\ref{2.30}) and curve T1 is the sum of the previous
contributions for the electron energy $\varepsilon=25$~GeV;
 curves 2, c2, T2 are the same the electron energy
 $\varepsilon=250$~GeV;
 curves 3, c3, T3 are the same for the electron energy
$\varepsilon=2.5$~TeV; curves 4, c4, T4 are the same for the
electron energy $\varepsilon=25$~TeV.
\label{fig.1}}
\end{figure}

 For relatively low energies
$\varepsilon=25~$GeV and $\varepsilon=8~$GeV used in famous SLAC
experiment \cite{12}, \cite{E2} we have analyzed the soft part of
spectrum, including all the accompanying effects: the boundary
photon emission, the multiphoton radiation and influence of the
polarization of the medium (see below). The perfect agreement of
the theory and data was achieved in the whole interval of measured
photon energies (200~keV$ \leq \omega \leq$500~MeV), see below
and the corresponding figures in \cite{BK1}, \cite{14}, \cite{15}.
It should be pointed out that both the correction term with
$F(\nu)$ and the Coulomb corrections have to be taken into
account for this agreement.

When a scattering is weak ($\nu_1 \ll 1$), the main contribution
in (\ref{2.30}) gives a region where $z \ll 1$. Then
\begin{eqnarray}
&&\displaystyle{-{\rm Im}~F(\nu)={\rm Im}~\nu^2\left(R_2- R_1
\right)/9,\quad L_c \rightarrow L_1},
\nonumber \\
&&\displaystyle{\Phi(\nu) \simeq \nu^2\left(R_1+2R_2 \right)/6,
\quad (|\nu| \ll 1)} \label{2.33}
\end{eqnarray}
Combining the results obtained in (\ref{2.33}) we obtain the
spectral distribution of the probability of radiation in the case
when scattering is weak $(|\nu| \ll 1)$
\begin{eqnarray}
\hspace{-6mm}&&\frac{dW}{d\omega}=\frac{dW_c}{d\omega}+\frac{dW_1}{d\omega}
=\frac{4Z^2\alpha^3n_a}{3m^2\omega}\Big[\frac{\omega^2}
{\varepsilon^2} \big(\ln \left(183Z^{-1/3} \right)
\nonumber \\
\hspace{-6mm}&&\displaystyle{-\frac{1}{6} -f(Z\alpha)\big)
+2\left(1+\frac{\varepsilon^{'2}}{\varepsilon^2} \right) \left(\ln
\left(183Z^{-1/3} \right)+\frac{1}{12}-f(Z\alpha) \right) \Big]},
\label{2.35}
\end{eqnarray}
where $L_1$ is defined in (\ref{2.18}). This expression coincide
with the known Bethe-Maximon formula for the probability of
bremsstrahlung from high-energy electrons in the case of complete
screening (if one neglects the contribution of atomic electrons)
written down within power accuracy (omitted terms are of the
order of powers of $\displaystyle{1/\gamma}$) with the Coulomb
corrections, see e.g. Eq.(18.30) in \cite{7}.

At $\nu_0 \gg 1$ the function ${\rm Im}~F(\nu)$ Eq.(\ref{2.30})
has the form
\begin{equation}
\displaystyle{-{\rm Im}~F(\nu)=\pi R_1/4+\nu_0 \left(\ln
2-C+\pi/4 \right)R_2/\sqrt{2}}. \label{2.38}
\end{equation}
Under the same conditions ($\nu_0 \gg 1$) the function ${\rm
Im}~\Phi(\nu)$ (\ref{2.25}) is
\begin{equation}
{\rm Im}~\Phi(\nu)=\pi R_1/4+\nu_0 R_2/\sqrt{2} \rightarrow \nu_0
R_2/\sqrt{2}.
\label{2.39}
\end{equation}
So, in the region where the LPM effect is strong the probability
(\ref{2.25}) can written as
\begin{equation}
\frac{dW_c}{d\omega}=\frac{\alpha R_2}{\varepsilon^2}
\left(\frac{Z^2\alpha^2 \varepsilon \varepsilon' n_a} {\pi
\omega} L(\varrho_c)\right)^{1/2}.
\label{2.39a}
\end{equation}
This means that in this limit the emission probability is
proportional to the square root of the density. This fact was
pointed out by Migdal \cite{M} (see Eq.(52)).

Thus, at $\nu_0 \gg 1$ the relative contribution of the first
correction $\displaystyle{\frac{dW_1}{d\omega}}$ is defined by
\begin{equation}
r=dW_1/dW_c= \left(\ln 2-C+\pi/4 \right)/2L(\varrho_c) \simeq
0.451/L(\varrho_c), \label{2.40}
\end{equation}
where $\displaystyle{L(\varrho_c)=\ln \frac{a_{s2}^2}{\lambda_c^2
\varrho_c^2}}$. In this expression the value $r$ with the
accuracy up to terms $\sim 1/L_1^2$ doesn't depend on the
energy:$L_c \simeq L_1+\ln(\varepsilon/\varepsilon_e)/2$. Hence
we can find the correction to the total probability at
$\varepsilon \gg \varepsilon_e$. The maximal value of the
correction is attained at $\varepsilon \sim 10\varepsilon_e$, it
is $\sim 6\%$ for heavy elements.

\section{Target of finite thickness}

\subsection{Boundary effects for a thick target}

For the homogeneous target of finite thickness $l$ the radiation
process in a medium depends on interrelation between $l$ and
formation length $l_{f0}$ (\ref{6}). In the case when $l \gg
l_{f0}$ we have the thick target where radiation on the boundary
should be incorporated. In the case when $l \ll l_{f0}$ we have
the thin target where the mechanism  of radiation is changed
essentially and in the case when $l \sim l_{f0}$ we have
intermediate thickness.

The spectral distribution of the probability of radiation of
boundary photons can be written in the form \cite{BK1}
\begin{equation}
\frac{dw_b}{d\omega}=\frac{4\alpha}{\omega} {\rm Re}\left<0|r_1
M+ r_2 {\bf p} M {\bf p}|0 \right>, \quad
 M=\left(\frac{1}{H+\kappa}-\frac{1}{{\bf p}^2+1} \right)^2,
\label{55}
\end{equation}
where $\kappa=1+\kappa_0^{'2},~\kappa_0^{'2}=
\kappa_0^2\varepsilon'/\varepsilon,~
\kappa_0=\omega_p/\omega,~\omega_p=\omega_0
\gamma,~\omega_0^2=4\pi \alpha n_e/m,~n_e$ is the electron
density, $\omega_0$ is the plasma frequency, $H$ is defined in
Eq.(\ref{2.9}),~$r_1=R_1\varepsilon'/\varepsilon=\omega^2/\varepsilon^2$,
~$r_2=R_2\varepsilon'/\varepsilon=1+\varepsilon^{'2}/\varepsilon^2$,
the parameter $\kappa_0$ describes the polarization of medium.

In the case when both the LPM effect and effect of polarization
of a medium are weak one can decompose combination in $M$ in
(\ref{55})
\begin{equation}
\frac{1}{H+\kappa}-\frac{1}{{\bf p}^2+1} \simeq \frac{1}{{\bf
p}^2+1}(iV-\kappa_0^{'2})\frac{1}{{\bf p}^2+1}. \label{51b}
\end{equation}
In the case when $\nu_1 \ll \kappa_0^2 \geq 1$ one can omit the
potential $V$ in $H$ (\ref{55}), so that the effect of
polarization of a medium is essential, then
\begin{eqnarray}
&&\frac{dw_b}{d\omega} =
\frac{4\alpha}{\omega(2\pi)^2}\int_{}^{}\left<0|r_1 M+ r_2 {\bf
p} M {\bf p}|0 \right>d^2p \nonumber \\
&& =\frac{\alpha}{\pi
\omega}\left\{r_1\left(1+\frac{1}{\kappa}-\frac{2}{\kappa-1}\ln
\kappa \right)+r_2\left[\left(1+\frac{2}{\kappa-1}\right)\ln
\kappa -2 \right]\right\} \label{51d}
\end{eqnarray}
This result is the quantum generalization of the transition
radiation probability.

\subsection{A thin target}

This is a situation when the formation length of radiation is much
larger than the thickness $l$ of a target \cite{14}
\begin{equation}
l \ll l_f =l_{f0}/\zeta,\quad \zeta = 1+\gamma^2\vartheta^2,\quad
T =l/l_{f0} \ll 1/\zeta,
\label{69}
\end{equation}
where $l_{f0}$ are defined in Eq.(\ref{6}). In the case $\kappa T
\ll 1$ the radiated photon is propagating in the vacuum and one
can neglect the polarization of a medium. The spectral
distribution of the probability of radiation from a thin target is
\begin{equation}
dw_{th}/d\omega=\alpha/\pi^2 \omega \int_{}^{} d^2\varrho
\left[r_1 K_0^2(\varrho)+r_2 K_1^2(\varrho) \right] \left(1-\exp
(-V(\varrho)T) \right),
\label{72}
\end{equation}
where $V(\varrho)$ is defined in (\ref{2.9}), (\ref{2.9a}),
$K_n$ is the modified Bessel function.

\subsection{Multiphoton effects in energy loss spectra}

It should be noted that in the experiments \cite{12},\cite{E2} the
summary energy of all photons radiated by a single electron is
measured. This means that besides mentioned above effects there
is an additional "calorimetric" effect due to the multiple photon
radiation. This effect is especially important in relatively
thick used targets. Since the energy losses spectrum of an
electron is actually measured, which is not coincide in this case
with the spectrum of photons radiated in a single interaction,
one have to consider the distribution function of electrons over
energy after passage of a target \cite{15}.

We consider the spectral distribution of the energy losses. After
summing over $n$ the probability of the successive radiation of
$n$ soft photons with energies  $\omega_1, \omega_2, \ldots
\omega_k$ by a particle with energy $\varepsilon~(\omega_k \ll
\varepsilon, k=1,2 \dots n)$ under condition $\sum_{k=1}^n
\omega_k=\omega$ on the length $l$ in the energy intervals
$d\omega_1 d\omega_2 \ldots d\omega_n$ we obtain \cite{15}
\begin{equation}
\frac{d\varepsilon}{d\omega} =\frac{1}{\pi}~{\rm Re}
\int_{0}^{\infty}\exp\left(ix \right)
\exp\left\{-\int_{0}^{\infty}\frac{dw}{d\omega_1}
\left[1-\exp\left(-ix\frac{\omega_1}{\omega}\right)\right]
d\omega_1\right\}dx \label{y5}
\end{equation}
The formula (\ref{y5}) was derived by Landau \cite{y9a} (see also
\cite{LP}) as solution of the kinetic equation under assumption
that energy losses are much smaller than particle's energy (the
paper \cite{y9a} was devoted to the ionization losses). The
energy losses are defined by the hard part of the radiation
spectrum. In the soft part of the energy losses spectrum
(\ref{y5}) the probability of radiation of one hard photon only
is taken into account accurately. So,it is applicable for the thin
targets only and has an accuracy $l/L_{rad}$.

We will analyze first the interval of photon energies where the
Bethe-Maximon formula is valid. Substituting this formula for
$dw/d\omega$ (within the logarithmic accuracy) into Eq.(\ref{y5})
we have
\begin{equation}
\frac{d\varepsilon}{d\omega}= \beta f_{BM},~ f_{BM}=\frac{\exp
(-\beta \mu)}{\Gamma(1+\beta)},~\beta=\frac{4l}{3L_{rad}},~\mu=\ln
\frac{\varepsilon}{\omega}-\frac{5}{8}+C, \label{y12}
\end{equation}
where $\Gamma(z)$ is the Euler gamma function. If we consider
radiation of the one soft photon, we have from (\ref{y5})
$d\varepsilon/d\omega=\beta$. Thus, the formula (\ref{y12}) gives
additional "reduction factor" $f_{BM}$ which characterizes the
distortion of the soft Bethe-Maximon spectrum due to multiple
photons radiation. The emission of accompanying photons with
energy much lesser or of the order of $\omega$ changes the
spectral distribution on quantity order of $\beta$. However, if
one photon with energy $\omega_r > \omega$ is emitted, at least,
then photon with energy $\omega$ is not registered at all in the
corresponding channel of the calorimeter. Since mean number of
photons with energy larger than $\omega$ is determined by the
expression
\begin{equation}
\sum_{n=0}^{\infty} n w_n =w_{\omega}=\int_{\omega}^{\varepsilon}
(dw/d\omega)~d\omega, \label{y13a}
\end{equation}
where $w_n$ is the probability of radiation of $n$ soft photons.
 So, when
radiation is described by the Bethe-Maximon formula the value
$w_{\omega}$ increases as a logarithm with $\omega$ decrease
($w_{\omega} \simeq \beta \ln \varepsilon/\omega$) and for large
ratio $\varepsilon/\omega$ the value $w_{\omega}$ is much larger
than $\beta$. Thus, amplification of the effect is connected with
a large interval of the integration ($\omega \div \varepsilon$) at
evaluation of the radiation probability.

If we want to improve accuracy of the formula (\ref{y5}) and for
the case of thick targets $l \geq L_{rad}$ one has to consider
radiation of an arbitrary number of hard photons. This problem is
solved in Appendix of \cite{15} for the case when hard part of the
radiation spectrum is described by the Bethe-Maximon formula. In
this case the formula (\ref{y5}) acquires the additional factor.
As a result we extend this formula on the case thick targets.

The Bethe-Maximon formula becomes inapplicable for the photon
energies $\omega \leq \omega_c$, where LPM effect starts to
manifest itself (see  Eq.(\ref{2.31b})). Calculating the integral
in (\ref{y5}) we find for the distribution of the spectral energy
losses
\begin{equation}
\frac{d\varepsilon}{d\omega}=3\beta
\sqrt{\frac{\omega}{2\omega_c}}f_{LPM}, \quad f_{LPM} =
\left(1+\frac{3\pi}{2\sqrt{2}}\beta
\sqrt{\frac{\omega}{\omega_c}} \right) \exp (-w_c), \label{y26}
\end{equation}
where $f_{LPM}$ is the reduction factor in the photon energy range
where the LPM effect is essential,
\begin{equation}
w_c=\beta\left(\ln \varepsilon/\omega_c+C_2 \right),\quad C_2
\simeq 1.959. \label{y21}
\end{equation}
In this expression the terms $\sim 1/L$ (see Sec.2) are not taken
into account.

In the region $\omega \ll \omega_p$ (see Eq.(\ref{55})) where the
contribution of boundary photons dominates the reduction factor
for transition radiation $f_{tr}$ was found in \cite{15}.

\subsection{Discussion of theory and experiment}

Here we compare the experimental data \cite{12},\cite{E2} with
theory predictions. According to Eq.(\ref{2.31b}) the LPM effect
becomes significant for $\omega \leq
\omega_c=\varepsilon^2/\varepsilon_e$. The mechanism of radiation
depends strongly on the thickness of the target. The thickness of
used target in terms of the formation length is
\begin{eqnarray}
&&\displaystyle{\beta(\omega) \equiv \frac{l}{l_{f}(\omega)}=
T\left(\nu_0+\kappa \right)\simeq T_c\left[
\frac{\omega}{\omega_c}+\sqrt{\frac{\omega}{\omega_c}}+
\frac{\omega_p^2}{\omega \omega_c} \right]},
\nonumber \\
&&\displaystyle{T=\frac{l \omega}{2\gamma^2},\quad
\omega_p=\omega_0 \gamma,\quad T_c \equiv T(\omega_c) \simeq
\frac{2\pi}{\alpha}\frac{l}{L_{rad}}}, \label{101}
\end{eqnarray}
where we put that $\displaystyle{\nu_0 \simeq
\sqrt{\frac{\omega_c}{\omega}}}$ (see Eq.(\ref{2.30b})). Below we
assume that $\omega_c \gg \omega_p$ which is true under the
experimental conditions. So we have that
\begin{equation}
T_c=2\pi l/\alpha L_{rad} \geq 20 \quad {\rm at} \quad l/L_{rad}
\geq 2~\%. \label{D1}
\end{equation}
For $T_c \gg 1$ ($\beta(\omega_c) \gg 1$) the minimal value of the
ratio of the thickness of a target to the formation length follows
from Eq.(\ref{101}):$\beta_m \simeq 2T_c
(\omega_p/\omega_c)^{2/3}$ and is attained for the heavy elements
(Au, W, U) at the initial energy $\varepsilon=25$~GeV. In this
case one has $\omega_c \simeq 250$~MeV, $\omega_p \simeq 4$~MeV,
$\beta_m \geq 2.5$.

\begin{figure}[ht]
\centerline{\epsfxsize=10cm\epsfbox{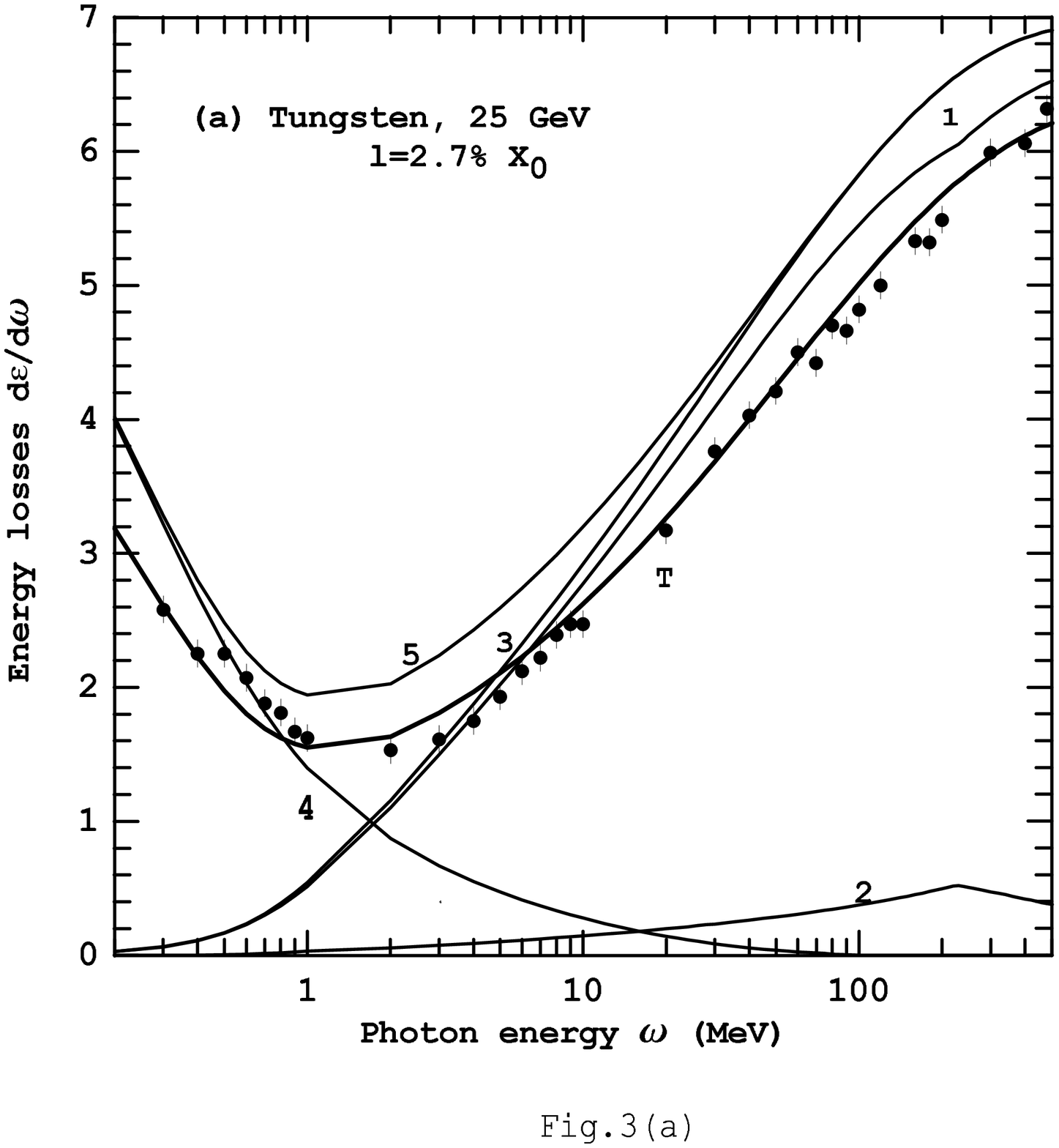}} \caption{The
energy losses $\omega dW/d\omega$ in gold with thickness
$l=0.023~mm$ in units $2\alpha/\pi$. Curve 1 is the main (Migdal)
contribution, curve 2 is the correction term, curve 3 is the sum
of previous contributions,
 curve 4 is the contribution of boundary photons,
 curve 5 is the sum of the previous contributions.
 curve T is the final theory prediction with regard to the
reduction factor (the multiphoton effects).
\label{fig.2}}
\end{figure}

As an example we calculated the spectrum of the energy losses in
the tungsten target with thickness $l=0.088$~mm (=$2.5~\%
L_{rad}$) for $\varepsilon$=25~GeV. The result is shown in Fig.2.
We calculated the main (Migdal type) term Eq.(\ref{2.25}), the
first correction term Eq.(\ref{2.30}) taking into account an
influence of the polarization of a medium, as well as the Coulomb
corrections entering into parameter $\nu_0$ Eq.(\ref{2.32a}) and
value $L(\varrho_c)$ Eq.(\ref{2.16}). We calculated also the
contribution of boundary photons (see Eq.(4.12) in \cite{BK1}).
Here in the soft part of the spectrum $\omega <
\omega_d~(\omega_d \simeq 2$ MeV ) the transition radiation term
Eq.(\ref{51d}) dominates, while in the harder part of the
boundary photon spectrum $\omega > \omega_d$ the terms depending
on both the multiple scattering and the polarization of a medium
give the contribution. All the mentioned contribution presented
separately in Fig.2. Under conditions of the experiment the
multiphoton reduction of the spectral curve is very essential. It
is taken into account in the curve "T". Experimental data are
taken from \cite{12}.  It is seen that there is a perfect
agreement of the curve T with data.

\section{Conclusion}

The LPM effect is essential at very high energies. It will be
important in electromagnetic calorimeters in TeV range. Another
possible application is air showers from the highest energy cosmic
rays. Let us find the lower bound of photon energy starting from
which the LPM effect will affect substantially the shower
development. The interaction of photon with energy lower than this
bound moving vertically towards earth surface will be described by
standard (Bethe-Maximon) formulas. The probability $w_p(h)$ to
find the primary photon on the altitude $h$ is
\begin{equation}
 dw_p/dh=w_p\exp(-h/h_0)/l_p,~w_p(h)=\exp\left(-h_0 e^{-h/h_0}/l_p\right),
 \label{C1}
\end{equation}
where $l_p \simeq 9l_r/7,~l_r$ is the radiation length on the
ground level ($l_r \simeq $ 0.3~km), $h_0 \simeq 8.7~$ km. The
parameter $\nu_p^2$ (see Eq.(\ref{2.30b})) characterizing the
maximal strength of the LPM effect at pair creation by a photon
$(\varepsilon_+=\varepsilon_-=\omega/2)$ has a form
\begin{equation}
\nu_p^2=(\omega/\omega_e)\exp(-h/h_0)w_p(h)
=(\omega/\omega_e)\exp(-h_0 e^{-h/h_0}/l_p-h/h_0),
 \label{C2}
\end{equation}
where $\omega_e$ is the characteristic critical energy for air on
the ground level ($\omega_e=4\varepsilon_e \simeq 10^{18}$~eV).
The last expression has a maximum at $h_m=h_0\ln(h_0/l_p)$, so
that $\nu_p^2(h_m)=\omega/\omega_m,~\omega_m=\omega_e h_0 e/l_p
\simeq 6\cdot 10^{19}$eV. Bearing in mind that the LPM effect
becomes essential for photon energy order of magnitude higher
than $\omega_m$ (e.g. the probability of pair creation is half of
Bethe-Maximon value at $\omega =12 \omega_e$) we have that the LPM
effect becomes essential for shower formation for photon energy
$\omega \simeq 5\cdot10^{20}$~eV. It should be noted that GZK
cutoff for proton is $5\cdot10^{19}$~eV. So the the photons with
the mentioned energy are extremely rare.


\begin{thebibliography}{99}
\bibitem{LP} L. D. Landau and I. Ya. Pomeranchuk, {\it Dokl.Akad.Nauk
SSSR} {\bf 92}, 535, 735 (1953). See in English in {\it The
Collected Papers of L. D. Landau}, Pergamon Press, 1965.
\bibitem{M} A. B. Migdal, {\it Phys. Rev.} {\bf 103}, 1811 (1956).
\bibitem{BKS1} V.N.Baier, V.M.Katkov and V.M.Strakhovenko,
{\it Sov. Phys. JETP} {\bf 67}, 70  (1988).
\bibitem{BKS2} V.N.Baier, V.M.Katkov and V.M.Strakhovenko,
{\it Electromagnetic Processes at High Energies in Oriented Single
Crystals}, World Scientific Publishing Co, Singapore, 1998.
\bibitem{BK1} V. N. Baier and V. M. Katkov,
{\it Phys.Rev.} {\bf D57}, 3146 (1998).
\bibitem{L5} V. N. Baier and V. M. Katkov,
{\it Phys.Rev.} {\bf D62}, 036008 (2000).
\bibitem{12} P. L. Anthony, R. Becker-Szendy, P. E. Bosted {\em et al},
{\it Phys.Rev}.{\bf D56}, 1373 (1997).
\bibitem{E2} S.Klein, {\it Rev. Mod. Phys.} {\bf 71}, 501 (1999).
\bibitem{14} V. N. Baier and V. M. Katkov,
{\it Quantum Aspects of Beam Physics}, ed.P. Chen, World
Scientific PC, Singapore, 1998, p.525.
\bibitem{15} V. N. Baier and V. M. Katkov,
{\it Phys.Rev.} {\bf D59}, 056003 (1999).
\bibitem{7} V. N. Baier, V. M. Katkov and V. S Fadin,
{\it Radiation from Relativistic Electrons} (in Russian)
Atomizdat, Moscow, 1973.
\bibitem{y9a} L. D. Landau J.Phys. USSR {\bf 8} (1944) 201.



\end{thebibliography}
\end{document}